\documentclass[a4paper, 10pt, conference]{IEEEtran}      

\IEEEoverridecommandlockouts                              

\usepackage[utf8]{inputenc}
\usepackage[T1]{fontenc}
\usepackage[utf8]{inputenc}
\usepackage{etex}
\usepackage{amsmath, amssymb, amsfonts, bm, nccmath}                        
\usepackage{booktabs}                       
\usepackage{graphicx}
\usepackage{amssymb, stfloats, amssymb, amsfonts, rotating, acronym}
\usepackage{multirow}
\usepackage{relsize}
\usepackage{array}
\usepackage{tabularx}
\usepackage{tikz}
\usetikzlibrary{patterns}
\usetikzlibrary{arrows,positioning,shapes.geometric,spy,matrix}
\usetikzlibrary{backgrounds}
\usepackage{mathtools}
\usepackage{color}
\usepackage{xcolor}
\usepackage{pgfplots}
\pgfplotsset{compat=newest}
\usepgfplotslibrary{groupplots}
\usepgfplotslibrary{fillbetween}
\usepackage{cite}
\usepackage{adjustbox}
\usepackage{colortbl}
\usepackage{lipsum}
\usepackage[linesnumbered,algoruled,boxed,lined]{algorithm2e}
\usepackage[normalem]{ulem}
\usepackage{cuted}

\def \absY {{|Y|}}
\def \absy {{|y_i|}}

\def \PJim {p_{S_{i-1}|Y_{i-1}}(J|y_{i-1})}
\def \PJip {p_{S_{i+1}|Y_{i+1}}(J|y_{i+1})}
\def \PJimA {p_{S_{i-1}|Y_{i-1}}(A|y_{i-1})}
\def \PJipA {p_{S_{i+1}|Y_{i+1}}(A|y_{i+1})}

\def \pest {\widehat{p_{S_i|Y^n}}(J|y^n)}

\newcommand{\fixme}[2]{\ifx&#2&{\leavevmode\color{red}#1}\else{\leavevmode\color{red}FIXME\{}#1{\leavevmode\color{red}\}}\footnote{{\leavevmode\color{red}#2}}\PackageWarning{Fixme}{#1: #2}\fi}

\definecolor{ocre}{RGB}{0, 157, 224} 
\definecolor{Paired-2}{RGB}{166,206,227}
\definecolor{Paired-1}{RGB}{31,120,180}
\definecolor{Paired-4}{RGB}{178,223,138}
\definecolor{Paired-3}{RGB}{51,160,44}
\definecolor{Paired-6}{RGB}{251,154,153}
\definecolor{Paired-5}{RGB}{227,26,28}
\definecolor{Paired-8}{RGB}{253,191,111}
\definecolor{Paired-7}{RGB}{255,127,0}
\definecolor{Paired-10}{RGB}{202,178,214}
\definecolor{Paired-9}{RGB}{106,61,154}
\definecolor{Paired-12}{RGB}{255,255,153}
\definecolor{Paired-11}{RGB}{177,89,40}
\definecolor{Accent-1}{RGB}{127,201,127}
\definecolor{Accent-2}{RGB}{190,174,212}
\definecolor{Accent-3}{RGB}{253,192,134}
\definecolor{Accent-4}{RGB}{255,255,153}
\definecolor{Accent-5}{RGB}{56,108,176}
\definecolor{Accent-6}{RGB}{240,2,127}
\definecolor{Accent-7}{RGB}{191,91,23}
\definecolor{Accent-8}{RGB}{102,102,102}
\definecolor{Spectral-1}{RGB}{158,1,66}
\definecolor{Spectral-2}{RGB}{213,62,79}
\definecolor{Spectral-3}{RGB}{244,109,67}
\definecolor{Spectral-4}{RGB}{253,174,97}
\definecolor{Spectral-5}{RGB}{254,224,139}
\definecolor{Spectral-6}{RGB}{255,255,191}
\definecolor{Spectral-7}{RGB}{230,245,152}
\definecolor{Spectral-8}{RGB}{171,221,164}
\definecolor{Spectral-9}{RGB}{102,194,165}
\definecolor{Spectral-10}{RGB}{50,136,189}
\definecolor{Spectral-11}{RGB}{94,79,162}
\definecolor{Set1-1}{RGB}{228,26,28}
\definecolor{Set1-2}{RGB}{55,126,184}
\definecolor{Set1-3}{RGB}{77,175,74}
\definecolor{Set1-4}{RGB}{152,78,163}
\definecolor{Set1-5}{RGB}{255,127,0}
\definecolor{Set1-6}{RGB}{255,255,51}
\definecolor{Set1-7}{RGB}{166,86,40}
\definecolor{Set1-8}{RGB}{247,129,191}
\definecolor{Set1-9}{RGB}{153,153,153}
\definecolor{Set2-1}{RGB}{102,194,165}
\definecolor{Set2-2}{RGB}{252,141,98}
\definecolor{Set2-3}{RGB}{141,160,203}
\definecolor{Set2-4}{RGB}{231,138,195}
\definecolor{Set2-5}{RGB}{166,216,84}
\definecolor{Set2-6}{RGB}{255,217,47}
\definecolor{Set2-7}{RGB}{229,196,148}
\definecolor{Set2-8}{RGB}{179,179,179}
\definecolor{Dark2-1}{RGB}{27,158,119}
\definecolor{Dark2-2}{RGB}{217,95,2}
\definecolor{Dark2-3}{RGB}{117,112,179}
\definecolor{Dark2-4}{RGB}{231,41,138}
\definecolor{Dark2-5}{RGB}{102,166,30}
\definecolor{Dark2-6}{RGB}{230,171,2}
\definecolor{Dark2-7}{RGB}{166,118,29}
\definecolor{Dark2-8}{RGB}{102,102,102}
\definecolor{Reds-1}{RGB}{255,245,240}
\definecolor{Reds-2}{RGB}{254,224,210}
\definecolor{Reds-3}{RGB}{252,187,161}
\definecolor{Reds-4}{RGB}{252,146,114}
\definecolor{Reds-5}{RGB}{251,106,74}
\definecolor{Reds-6}{RGB}{239,59,44}
\definecolor{Reds-7}{RGB}{203,24,29}
\definecolor{Reds-8}{RGB}{165,15,21}
\definecolor{Reds-9}{RGB}{103,0,13}
\definecolor{Greens-1}{RGB}{247,252,245}
\definecolor{Greens-2}{RGB}{229,245,224}
\definecolor{Greens-3}{RGB}{199,233,192}
\definecolor{Greens-4}{RGB}{161,217,155}
\definecolor{Greens-5}{RGB}{116,196,118}
\definecolor{Greens-6}{RGB}{65,171,93}
\definecolor{Greens-7}{RGB}{35,139,69}
\definecolor{Greens-8}{RGB}{0,109,44}
\definecolor{Greens-9}{RGB}{0,68,27}
\definecolor{Blues-1}{RGB}{247,251,255}
\definecolor{Blues-2}{RGB}{222,235,247}
\definecolor{Blues-3}{RGB}{198,219,239}
\definecolor{Blues-4}{RGB}{158,202,225}
\definecolor{Blues-5}{RGB}{107,174,214}
\definecolor{Blues-6}{RGB}{66,146,198}
\definecolor{Blues-7}{RGB}{33,113,181}
\definecolor{Blues-8}{RGB}{8,81,156}
\definecolor{Blues-9}{RGB}{8,48,107}
\definecolor{awesome-skyblue}{HTML}{0395DE}
\definecolor{bg}{HTML}{282828}

\definecolor{rosso}{RGB}{220,57,18}
\definecolor{giallo}{RGB}{255,153,0}
\definecolor{blu}{RGB}{102,140,217}
\definecolor{verde}{RGB}{16,150,24}
\definecolor{viola}{RGB}{153,0,153}

\definecolor{bleuUni}{RGB}{0, 157, 224}
\definecolor{marronUni}{RGB}{68, 58, 49}

\title{A General Security Approach for Soft-information Decoding against Smart Bursty Jammers}

\begin{document}

\pgfmathdeclarefunction{gauss}{2}{%
  \pgfmathparse{1/(#2*sqrt(2*pi))*exp(-((x-#1)^2)/(2*#2^2))}%
}

\author{\IEEEauthorblockN{Furkan Ercan\textsuperscript{$\dagger$}, Kevin Galligan\textsuperscript{*}, Ken R. Duffy\textsuperscript{*}, Muriel M{\'e}dard\textsuperscript{$\mathsection$}, David Starobinski\textsuperscript{$\dagger$}, Rabia Tugce Yazicigil\textsuperscript{$\dagger$}}
\IEEEauthorblockA{\textsuperscript{$\dagger$}Department of Electrical and Computer Engineering, Boston University, Boston, MA, USA \\
\textsuperscript{$\mathsection$}Department of Electrical Engineering and Computer Science, MIT, Cambridge, MA, USA \\
\textsuperscript{*}Hamilton Institute, Maynooth University, Ireland
}}

\maketitle

\thispagestyle{empty}
\pagestyle{empty}

\begin{abstract}

Malicious attacks such as jamming can cause significant disruption or complete denial of service (DoS) to wireless communication protocols. Moreover, jamming devices are getting smarter, making them difficult to detect. Forward error correction, which adds redundancy to data, is commonly deployed to protect communications against the deleterious effects of channel noise. Soft-information error correction decoders obtain reliability information from the receiver to inform their decoding, but in the presence of a jammer such information is misleading and results in degraded error correction performance. As decoders assume noise occurs independently to each bit, a bursty jammer will lead to greater degradation in performance than a non-bursty one. Here we establish, however, that such temporal dependencies can aid inferences on which bits have been subjected to jamming, thus enabling counter-measures. In particular, we introduce a pre-decoding processing step that updates log-likelihood ratio (LLR) reliability information to reflect inferences in the presence of a jammer, enabling improved decoding performance for any soft detection decoder.
The proposed method requires no alteration to the decoding algorithm. Simulation results show that the method correctly infers a significant proportion of jamming in any received frame. Results with one particular decoding algorithm, the recently introduced ORBGRAND, show that the proposed method reduces the block-error rate (BLER) by an order of magnitude for a selection of codes, and prevents complete DoS at the receiver.

\end{abstract}

\section{Introduction}


Jammers typically aim to cause a denial of service (DoS) or reduction of quality (RoQ) at the receiver \cite{orakcal2014jamming} without getting detected. They exploit the wireless transmission by mixing their signals with legitimate communication. As a result, the received frame becomes undecodable, which causes anomalies such as increased repeat requests, reduced throughput, prolonged delays, or a complete breakdown \cite{dos-attacks-ieeecomm-2011}. Powerful jammers that blast channels with unrestrained amounts of energy can be detected easily by the receiver. More subtle jammers, on the other hand, might seek to inject short bursts or lower levels of energy to disrupt communication while circumventing their detection, causing a DoS. In general, jammers must demonstrate high energy efficiency, low detection probability, high levels of DoS, and resistance against physical layer (PHY) anti-jamming techniques.

From an information-theoretic perspective, uniform jammers are the most effective for reducing the channel capacity and the code rate \cite{turner1979}. However, emerging techniques such as rate-adaptation algorithms propose efficient countermeasures for such jammer attacks \cite{Gawas2017}. On the other hand, bursty jammers \cite{orackal2012} can be an effective approach for increasing the block-error rate (BLER), where an adversary jams a burst of bits in a transmitted frame. Bursty jammers become more effective in increasing the BLER when their burst patterns are unpredictable to the receiver. With increased BLER, the receiver must compensate by reducing the code rate, which sacrifices information throughput. Therefore it is essential to study countermeasures to such jammer attacks.

Most traditional security approaches for wireless technologies are applied to upper layers in the protocol stack \cite{mukherjee2014principles}. However, with the rapid growth in use cases and network density, maintaining security for 5G-and-beyond technologies has become a challenge \cite{Ghasempour2022}. PHY-layer security is an emerging solution to threats that arise with evolving adversaries \cite{wu2018survey}. Under such adversarial behavior, machine learning-based approaches \cite{yuxin2021,upad2019} and spectrum sensing-based approaches \cite{upad2021} have been proposed to counter jamming.
Our paper specifically focuses on jamming attacks on soft-information decoders, a topic that has received scant attention in the literature. Our anti-jamming approach applies to general coding schemes and can be effortlessly supported on the physical layer with minimal computational overhead.

In this work, we consider a smart, reactive jammer that is bursty and only active during a fraction of the transmission. It is assumed that  transmission parameters, such as the modulation and the subcarrier frequency, are known to the adversary. To counter such an attack, we propose a modified log-likelihood ratio (LLR) computation that takes the conditional probability of jamming into account for each index of the received frame. The computation of this posterior probability is performed in two steps. First, an initial value is calculated based on the received signal strength. Anchor points in the received frame, for which the conditional jamming probability is high, are then used to inform the jamming estimates of neighbouring points, based on Markov state transition probabilities. The proposed method is general to any receiver and carried out before decoding. Simulation results show that the proposed method unveils a significant amount of the attack, and therefore the attacker cannot maintain their deniability. Using the universal ORBGRAND algorithm \cite{duffy2021orbgrand,duffy2022orbgrand}, it is shown that an order of magnitude of BLER performance can be recovered with the proposed method and a complete DoS is prevented, using different codebooks, \textit{i.e.} random linear codes (RLCs) and 5G cyclic redundancy check-aided Polar codes (CA-Polar).

The rest of the paper is organized as follows. In Section~\ref{sec:bg}, preliminaries are detailed. In Section~\ref{sec:llr}, the smart bursty jammer model and proposed LLR approach with the conditional jamming probability computation is presented. Section~\ref{sec:pj} explains how to approximate the conditional probability of jamming. Results are presented in Section~\ref{sec:res}, followed by concluding remarks in Section~\ref{sec:conc}.

\section{Preliminaries}\label{sec:bg}

\subsection{PHY Jammer Models}\label{sec:bg:jammer}

Protection against an adversary is not possible if the adversary has unlimited resources. Hence, we assume that the adversary must operate under a set of constraints. 
A fully modeled adversary must have assumptions, goals, and capabilities \cite{do2019role}. Although there are numerous categorizations of jammers in the literature, the PHY jammer models can be summarized in the following two categories \cite{dos-attacks-ieeecomm-2011,xu2005feasibility}. 

\subsubsection{Constant jammers}
As their name suggests, constant jammers continuously emit disruptive signals over the communication medium. Constant jammers are primitive and often can be detected through the radio signal strength indicator (RSSI) component of the receiver. Simple measures such as frequency hopping can be taken as a precaution against these types of jammers \cite{rty-freqhopping}. Moreover, constant jammers are power inefficient, which limits their ability to be mobile. 
\subsubsection{Reactive jammers}
As a power-efficient and more intelligent alternative, reactive jammers emit signals only when it senses a legitimate transmission taking place. This type of jammer causes a signal collision at the receiver that disrupts either part of or all of the frame. Prevention techniques for these types of jammers include interference and RSS sampling \cite{strasser2010detection}. Carefully engineered, smart, reactive jammers are the most challenging type of jammer \cite{huynh2020jammer}.

Usually, the error correction algorithms embedded in the PHY can be considered as a first response against such undesired attacks. However, as the error-correcting codes (ECCs) are standardized, their error correction capability is known to the adversary. Therefore, a jammer can corrupt just enough amount of transmission to cause  the decoding to fail, eventually causing a DoS.

\subsection{Channel model}\label{sec:bg:llr}
Every soft-information decoder requires LLR as an input which determines the hard output value of each received signal, and also acts as a measure of \textit{reliability} for those signals. In regular conditions, a larger LLR magnitude indicates more confidence in the received signal. 

Let $b^n$,  a binary channel input of length $n$, be modulated using binary phase-shift keying (BPSK) with the mapping
\begin{equation*}
b^n \in \{0, 1\}^n \rightarrow x^n \in \{+1, -1\}^n,
\end{equation*}
where $x^n$ is the modulated channel input variable sequence. 
Assuming equiprobable symbols and IID noise, given a realization of the received signal, $y^n = (y_0, y_1, \cdots, y_{n-1})$, 
the LLRs can be calculated per-bit as
\begin{equation}\label{eqn:llr:awgn}
    L(y_i|A) = \frac{2y_i}{\sigma_A^2} \text{, for each }  i \in \{1,\ldots,n\},
\end{equation}
where $i$ indicates the bit index of the received frame, the conditioning on $A$ indicates it is an AWGN channel without jamming, and $\sigma_A$ is the standard deviation of the channel noise.

\section{Evaluating LLRs Under Jamming}\label{sec:llr}

\subsection{Threat Model}

The adversary is modeled as a jammer which disguises itself by injecting zero-mean Gaussian noise into the system. It is assumed that the smart jammer can retrieve the modulation and subcarrier frequency of operation and therefore injects jammer signals at the legitimate transmission frequency. In order not to alert RSSI of the transmission system, the jammer interferes only a fraction of the time and does so randomly in a bursty fashion. The occurrence of jamming is modeled as a Markov chain at the level of transmitted bits.

\begin{figure}
\includegraphics[width=\columnwidth]{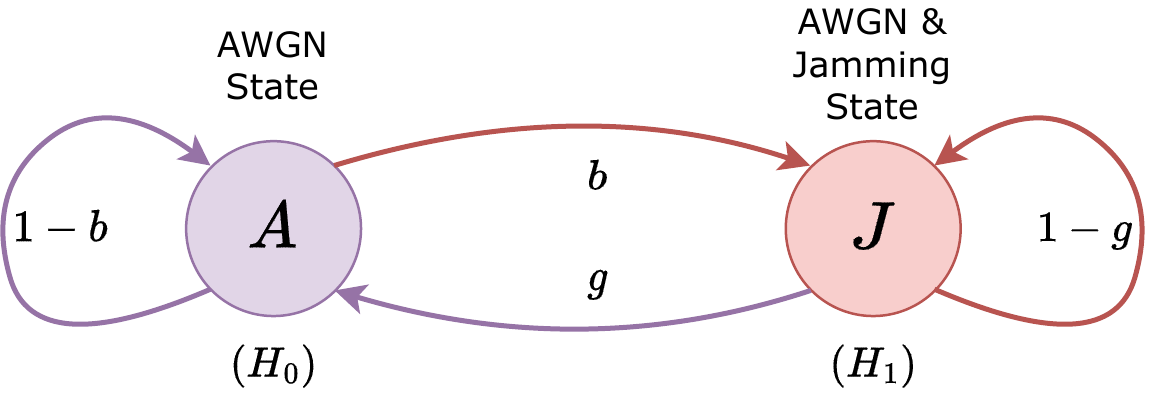}
\caption{Two-state Markov chain model for the reactive jammer model, with transition probabilities $b$ and $g$. The state of the chain for bit $i$ is denoted $S_i$.} 
\vspace{-1em}
\label{fig:markovchain}
\end{figure}

Fig.~\ref{fig:markovchain} depicts the two-state Markov chain for the jammed channel model. The state $A$ is AWGN only with zero mean and variance $\sigma^2_{A}$. The $J$ state denotes that jamming is present in the channel, with total variance $\sigma^2_{J}$:
\begin{equation}\label{eqn:sigma_j}
    \sigma^2_{J} = \sigma^2_{V} + \sigma^2_{A} \text{.}
\end{equation}
Here, $\sigma^2_{V}$ is the variance of the signal introduced by the jammer, which is an independent Gaussian random variable. The state transitions are modeled to occur per-bit. The state transition parameters $b$ and $g$ denote the probabilities of passing from the AWGN state to the jamming state and vice versa, respectively. The parameters $b$, $g$, $\sigma^2_{J}$, and $\sigma^2_{A}$ can be estimated, and so are assumed known to the receiver. 

\subsection{LLR Calculation Under Jamming}
Given that a received signal $y_i$ is certainly affected by jamming, then its noise is independent from that impacting other
bits and the LLR would be
\begin{equation}\label{eqn:llr:jam}
    L(y_i|J) = \frac{2y_i}{\sigma_J^2} 
\end{equation}
instead of (\ref{eqn:llr:awgn}), where $\sigma_J^2$ is obtained using (\ref{eqn:sigma_j}). In practice, however, the receiver does not have certainty on whether a signal has been impacted by jamming and that induces hidden Markov dependencies in the calculation of the LLRs. Regardless, the decoder will treat the LLR of each bit as being an independent random variable and so the objective is to provide the best marginal estimate of the LLR of each bit given the jamming uncertainty.

Let $\{S_i\}$ denote the Markov state process, with $S_i$ taking values of $A$ for the AWGN state and $J$ for the jamming state. Then, the conditional probability of the transmitted binary variable $B_i$ at index $i$ being a $0$ can be computed as
\begin{align}
    p_{B_i|Y^n}(0|y^n) & = \sum_{s^n\in\{A,J\}^n} p_{B_i,S^n|Y^n}(0,s^n|y^n) \notag \\
        & = \sum_{s^n\in\{A,J\}^n} p_{B_i|S^n,Y^n}(0|s^n,y^n) p_{S^n|Y^n}(s^n|y^n)
    \label{eq:bpost}
\end{align}
taking the entire received signal into account and accordingly, its marginal LLR would be
\begin{equation}
    L(y_i) =  \ln \frac{p_{B_i|Y^n}(0|y^n)}{p_{B_i|Y^n}(1|y^n)}
\end{equation}
which can be expanded to incorporate the jamming uncertainty using equation \eqref{eq:bpost}.

Given the received signal sequence $y^n$, the conditional probability of a jamming sequence $s^n\in\{A,J\}^n$ can be computed as
\vspace{-0.5em}
\begin{equation}\label{eqn:pjx}
   p_{S^n|Y^n}(s^n|y^n) = \frac{f_{Y^n|S^n}(y^n|s^n) p_{S^n}(s^n)}{f_{Y^n}(y^n)}.
\end{equation}
where $f$ is the probability density function (PDF). As the noise is independent of the channel states, we have that
\begin{equation}\label{eqn:pj_pmf}
    f_{Y^n|S^n}(y^n|s^n) = \prod_{i=1}^n f_{Y|S}(y_i|s_i) \text{.}
\end{equation}
Incorporating (\ref{eqn:pj_pmf}) into (\ref{eqn:pjx}), we get
\begin{equation}\label{eqn:pjx_new}
    p_{S^n|Y^n}(s^n|y^n) = \frac{\prod_{i=1}^n f_{Y|S}(y_i|s_i) p_{S^n}(s^n) }{ f_{Y^n}(y^n) }\text{,}
\end{equation}
where $s^n$ ranges over $2^n$ possible jamming sequences. 
The probability of a received signal at an arbitrary index $i$ being in the $J$ state can be evaluated from \eqref{eqn:pjx} as
\begin{equation}\label{eqn:bruteforce}
p_{S_i|Y^n}(J|y^n) = \sum_{s^n\in\{A,J\}^n:s_i=J} p_{S^n|Y^n}(s^n|y^n).
\end{equation}
The brute force evaluation in \eqref{eqn:bruteforce} requires a burdensome $2^{n-1}$ computations, so in the following section we propose an efficient estimation technique for the marginal probability of jamming. Moreover, for reduced computation, we employ a linear approximation to the full LLR computation unconditioned
on jamming state:
\vspace{-0.5em}
\begin{equation}\label{eqn:llr:proposed}
    \hat{L}(y_i) = L(y_i|A) p_{S_i|Y^n}(A|y^n) + L(y_i|J) p_{S_i|Y^n}(J|y^n).
\end{equation}

\section{Approximating the Conditional Probability of Jamming}\label{sec:pj}

\subsection{The Impact of False Positives/Negatives on BLER}\label{sec:pj:errortypes}

The collected statistical data, which is the received signal in our case, may lead to incorrect conclusions in terms of misidentifying the $A$ and $J$ states. Therefore, it is essential to assess the impact of false positives and false negatives on the BLER performance. 

\begin{figure}
\centering
\begin{tikzpicture}[spy using outlines=
	{circle, magnification=2.0, connect spies}]
  \pgfplotsset{
    label style = {font=\fontsize{9pt}{7.2}\selectfont},
    tick label style = {font=\fontsize{7pt}{7.2}\selectfont}
  }

\begin{axis}[
	scale = 1,
    ymode=log,
    xlabel={$\%$ False Positives}, xlabel style={yshift=0.4em},
    xticklabel={\pgfmathprintnumber\tick\%},
    ylabel={Expected BLER}, ylabel style={yshift=-0.50em},
    grid=both,
    ymajorgrids=true,
    xmajorgrids=true,
    grid style=dashed,
    width=0.55\columnwidth, height=7.0cm,
    thick,
    mark size=3,
    legend pos=north east,
    legend style={
    at={(2.00,0.25)},
     anchor={west},
     cells={anchor=west},
     column sep= 2mm,
      font=\fontsize{7.5pt}{7.2}\selectfont,
    },
    legend cell align={left},
    legend columns=2,
]


\addplot[
    color=Paired-1,
    mark=triangle,
    thick,
    mark size=3,
    mark repeat=2,
]
table {
0	2.88E-04
5	5.91E-04
10	9.08E-04
15	1.73E-03
20	2.43E-03
25	3.59E-03
30	5.10E-03
35	8.30E-03
40	1.12E-02
45	1.55E-02
50	1.89E-02
55	2.30E-02
60	2.92E-02
65	3.41E-02
70	4.19E-02
75	5.07E-02
80	6.19E-02
85	7.32E-02
90	8.20E-02
95	8.74E-02
100	9.21E-02

};

\end{axis}

\node at (1.7,-1.05) {(a)};

\end{tikzpicture}
\begin{tikzpicture}[spy using outlines=
	{circle, magnification=2.0, connect spies}]
  \pgfplotsset{
    label style = {font=\fontsize{9pt}{7.2}\selectfont},
    tick label style = {font=\fontsize{7pt}{7.2}\selectfont}
  }

\begin{axis}[
	scale = 1,
    ymode=log,
    xlabel={$\%$ False Negatives}, xlabel style={yshift=0.4em},
    xticklabel={\pgfmathprintnumber\tick\%},
    yticklabel=\empty,
    grid=both,
    ymajorgrids=true,
    xmajorgrids=true,
    grid style=dashed,
    width=0.55\columnwidth, height=7.0cm,
    thick,
    mark size=3,
    legend pos=north east,
    legend style={
    at={(2.00,0.25)},
     anchor={west},
     cells={anchor=west},
     column sep= 2mm,
      font=\fontsize{7.5pt}{7.2}\selectfont,
    },
    legend cell align={left},
    legend columns=2,
]


\addplot[
    color=Paired-1,
    only marks,
    mark=o,
    thick,
    mark size=3,
    mark repeat=2,
]
table {
0	2.88E-04
5	5.30E-03
10	1.07E-02
15	1.46E-02
20	2.01E-02
25	2.54E-02
30	2.99E-02
35	3.44E-02
40	3.90E-02
45	4.46E-02
50	4.89E-02
55	5.37E-02
60	5.95E-02
65	6.40E-02
70	6.85E-02
75	7.46E-02
80	7.81E-02
85	8.15E-02
90	8.63E-02
95	8.93E-02
100	9.21E-02

};

\addplot[
    color=Paired-1,
    thick,
]
table {
0	2.88E-04
1	1.34E-03
2	2.54E-03
3	3.52E-03
4	4.11E-03
5	5.30E-03
10	1.07E-02
15	1.46E-02
20	2.01E-02
25	2.54E-02
30	2.99E-02
35	3.44E-02
40	3.90E-02
45	4.46E-02
50	4.89E-02
55	5.37E-02
60	5.95E-02
65	6.40E-02
70	6.85E-02
75	7.46E-02
80	7.81E-02
85	8.15E-02
90	8.63E-02
95	8.93E-02
100	9.21E-02

};

\end{axis}

\node at (1.7,-1.05) {(b)};
\end{tikzpicture}
\vspace{-1em}
\caption{The quantified impact of (a) false positives and (b) false negatives on the BLER performance, using $\text{RLC}[128,105]$ with the ORBGRAND algorithm.}
\label{fig:type12}
\vspace{-1em}
\end{figure}
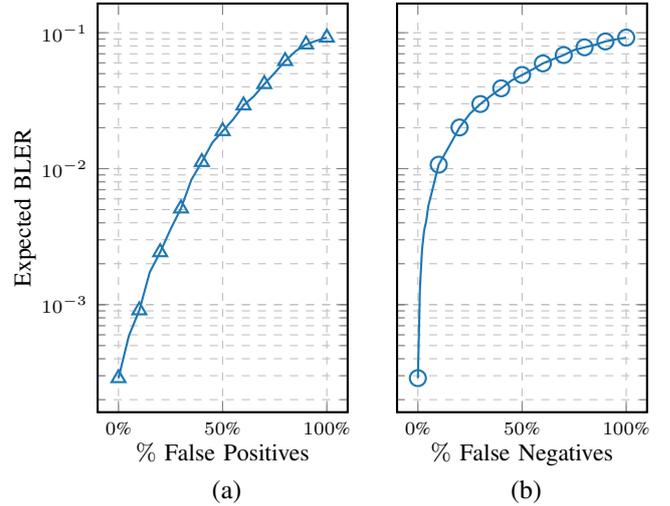

False positives occur when a non-jammed index is mistaken for being jammed. In this scenario, $L(y_i|J)$ in equation \eqref{eqn:llr:jam} is used instead of $L(y_i|A)$ in equation \eqref{eqn:llr:awgn} for the mistaken index $i$.
False negatives occur when a jammed index is mistaken for being non-jammed and  $L(y_i|A)$ is used instead of $L(y_i|J)$ for the mistaken index $i$. 

To understand and quantify the impact of mistaking the events on the BLER performance, a set of genie-aided simulations is carried out. A random linear code $\text{RLC}[n, k] = \text{RLC}[128, 105]$ is used as an example where $n$ denotes the code length and $k$ denotes the code dimension, and the universal ORBGRAND algorithm is used to derive the BLER performance. The state information for each received bit is provided to the genie-aided decoder, therefore, $L(y_i|A)$ is used for indices belonging to state $A$, and $L(y_i|J)$ is used otherwise. To quantify the impact of false positives, BLER is measured when $L(y_i|J)$ is used for a proportion of indices that belong to state $A$. Similarly, to quantify the impact of false negatives, BLER is measured when $L(y_i|A)$ is used for a proportion of indices that belong to state $J$. 

Fig.~\ref{fig:type12} presents the simulated BLER performance for the percentage of false positives (a) and false negatives (b). The SNRs for the AWGN channel and the jammer are selected as $\text{SNR}_A = 12$ and $\text{SNR}_J = 0$ dB, respectively. In both performance assessments, it can be observed that the BLER performance degrades as the number of errors increases. However, the degradation with false negatives is far more severe than the degradation with false positives. For instance, $5\%$ of false negatives has the same amount of impact on BLER performance as about $40\%$ of false positives. This means that the correct identification of jammed indices is far more important than the incorrect identification of the non-jammed indices, and our algorithm should prioritize identifying jammed indices correctly.

\subsection{Calculating the Jamming Probability}\label{sec:pj:calc1}

The estimation of probability of jamming is performed in two steps. In the first step, an initial estimate of the probability that the $i$-th bit experienced jamming, $p_{S_i|Y^n}(J|y^n)$, is derived based on the marginal distribution given $y_i$ alone $p_{S_i|Y_i}(J|y_i)$. Then, using the Markov state transition probabilities, the probability of jamming for specific indices neighboring those with high jamming likelihoods are recomputed to improve the estimates of their probabilities.

The sign of a received signal $y_i$ does not have an impact on $p_{S_i|Y_i}(J|y_i)$. Hence, we consider a new random variable, $\absY$, that is based on the magnitude of $Y$. In this case, the new random variable is a folded Gaussian distribution with  PDF, $f_{\absY}(\absy)$, equal to
\begin{equation}\label{eqn:foldedGauss}
\frac{1}{\sigma\sqrt{2\pi}} \bigg{(} \exp\Big{(}\frac{-(\absy-1)^2}{2\sigma^2}\Big{)} + \exp\Big{(}\frac{-(\absy+1)^2}{2\sigma^2}\Big{)} \bigg{)}
\end{equation}
for $0 \leq i < n$. In the first step, our estimate of $p_{S_i|Y^n}(J|y^n)$ is 
\begin{equation}\label{eqn:pj_init}
p_{S_i|Y_i}(J|y_i) = \frac{f_{\absY|S_i}(\absy\big{|}J)  p_{S_i}(J)}{f_{\absY}(\absy)}.
\end{equation}
The conditional PDF expression in (\ref{eqn:pj_init}) can be obtained by substituting the jamming variance in the expression in (\ref{eqn:foldedGauss}). Using the law of total probability, the PDF at the denominator in (\ref{eqn:pj_init}) is expanded as
\begin{equation}\label{eqn:bayes1:end}
f_{\absY}(\absy) = \sum_{s_i\in\{A,J\}} f_{\absY|S_i}(\absy|s_i) p_{S_i}(s_i) \text{.}
\end{equation}
Substituting (\ref{eqn:foldedGauss}) and (\ref{eqn:bayes1:end}) into (\ref{eqn:pj_init}), the first approximation for the conditional probability of bit $i$ having experienced jamming can be calculated.

\begin{figure}
  \centering
  \input{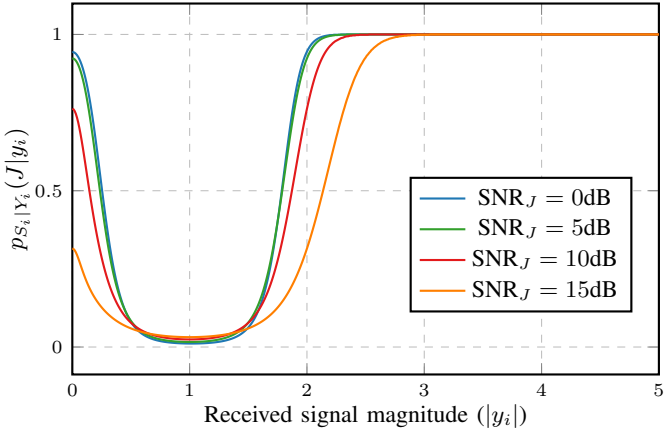}
  \caption{$p_{S_i|Y_i}(J|y_i)$ as a function of received signal magnitude, $|y_i|$, based on (\ref{eqn:pj_init}). The SNR of the AWGN channel is fixed at $\text{SNR}_A=12$ dB, and several probabilities are depicted based on various jamming SNRs.}
  \label{fig:pj}
\end{figure}

Fig.~\ref{fig:pj} presents $p_{S_i|Y_i}(J|y_i)$ as a function of the received signal magnitude $\absy$. It is minimized at the absolute value of the BPSK constellation point, $1$, and is maximized as the received signal magnitude drifts away from the constellation. Note that the $p_{S_i|Y_i}(J|y_i)$ takes the stationary probability of jamming at the constellation point since there is always a chance that the received signal could be a result of jamming. 

\begin{figure}
  \centering
  \input{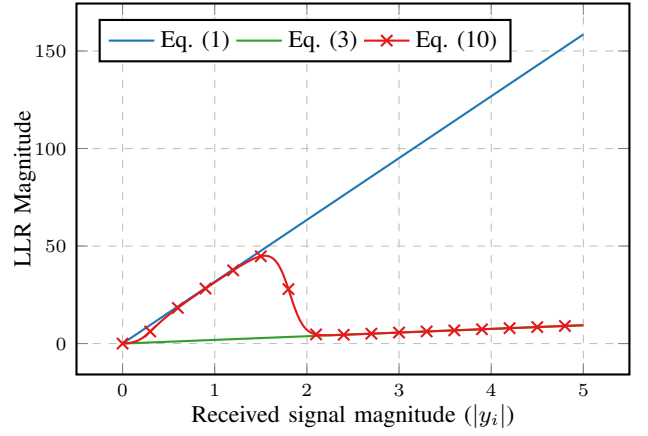}
  \caption{LLR magnitudes based on AWGN only (\ref{eqn:llr:awgn}), jamming only (\ref{eqn:llr:jam}), and proposed approach (\ref{eqn:llr:proposed}) using the first approximation to marginal conditional jamming probabilities. The SNRs of the AWGN channel and the jamming channel are fixed at 12 dB and 0 dB, respectively.}
  \label{fig:llr_new}
\end{figure}

Fig.~\ref{fig:llr_new} depicts LLR magnitude trend lines based on AWGN and jamming conditions, as well as the proposed LLR computation (\ref{eqn:llr:proposed}) when the first approximation $p_{S_i|Y_i}(J|y_i)$ (\ref{eqn:pj_init}) is incorporated. With increasing signal magnitude, the proposed method switches from the AWGN LLR trend line toward the jamming LLR trend line. This behavior reduces the strength of the LLRs at higher magnitudes as a result of the suspicion of jamming, which is then evaluated at soft-information decoders as a less reliable bit index. Consequently, such indices are naturally prioritized for correction, in attempts to identify the transmitted codeword.

When the jammer yields signal magnitude that is great enough to come under suspicion $p_{S_i|Y_i}(J|y_i)$ is a good estimate of $p_{S_i|Y^n}(J|y^n)$, as demonstrated in Fig.~\ref{fig:pj} and Fig.~\ref{fig:llr_new}. On the other hand, solely relying on the signal magnitudes would not allow us to detect a substantial portion of the jammed indices as indices with signal magnitudes close to the constellation point would mostly be inferred to be as non-jammed, which is a major limiting factor on the performance improvement. 

\begin{figure}
\includegraphics[width=\columnwidth]{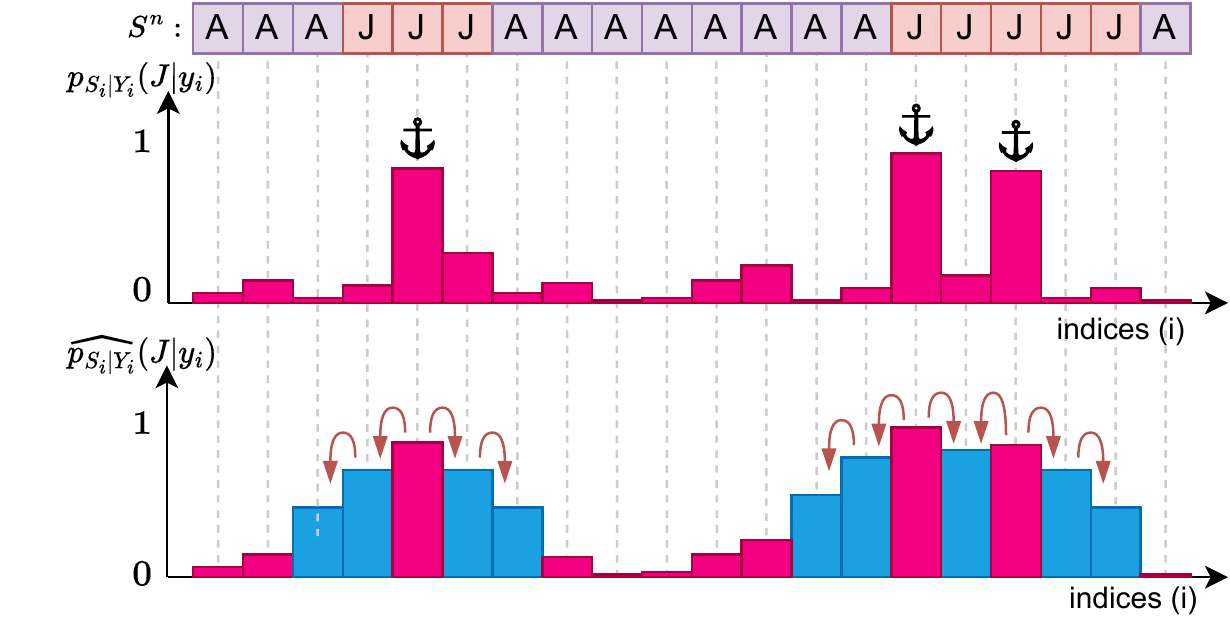}
\caption{Example state transition probability and their associated $p_{S_i|Y_i}(J|y_i)$. Indices with high $p_{S_i|Y_i}(J|y_i)$ values are designated as anchor indices (represented with the anchor symbol) and Markov chain state transition probabilities are used to recalculate $p_{S_i|Y_i}(J|y_i)$ for the neighboring indices resulting in a better estimate, $\pest$.}
\label{fig:anchor}
\end{figure}

To tackle this issue, we take advantage of the burstiness of the two-state Markov chain. If an index $i$ has a low initial $p_{S_i|Y_i}(J|y_i)$ value, but is neighboring an index $i \mp 1$ that has sufficiently high value, as governed by a threshold, then our estimate of $p_{S_i|Y_i}(J|y_i)$ is increased using a heuristic. This is illustrated in Fig.~\ref{fig:anchor} for a sequence of signals. On the top, the sequence $S^n$ represents the Markov state of a series of indices and is hidden from the receiver. The receiver calculates $p_{S_i|Y_i}(J|y_i)$, from which it determines a subset of indices that have a relatively high values. The indices at which $p_{S_i|Y_i}(J|y_i)$ yields a significantly high value are called \textit{anchor indices}. Using the Markov chain state transition probabilities, the $p_{S_i|Y_i}(J|y_i)$ for the indices adjacent to these anchor indices can be recalculated recursively. As a result, we derive a new, improved set of jamming probability estimations, $\pest$ for $i\in\{0,\ldots,n-1\}$.

In order to reconsider the jamming probability of an index, it must either be neighboring to an anchor index or be sandwiched between two distinct anchor indices. Otherwise, the initial $p_{S_i|Y_i}(J|y_i)$ is used.

\subsubsection{Index Neighboring to a Single Anchor Index}
In the first case, the index of interest neighbors an anchor index on one side and a non-anchor index on the other. For simplicity, let us consider the subject index $i$ and the anchor index $i-1$. Using the Markov property, we create an updated $\pest$ from its anchoring neighbour. Assuming the anchor is in the $i-1$ position, using the Markov property we set
\begin{align}\label{eqn:singleanchor}
    &\pest = \notag \\ 
    &b\PJimA+(1-g)\PJim \text{.}
\end{align}

\subsubsection{Index Neighboring to Two Anchor Indices}

Similar to (\ref{eqn:singleanchor}), we derive the updated jamming probability for an index that is in between two anchor indices. For the subject index located at $i$, the anchor indices are at $i-1$ and $i+1$. Unlike the previous case, the new probability is conditioned on two different states. Based on the values of $\PJim$, $\PJip$, $b$ and $g$ values, again using the Markov property $\pest$ is expressed as:
\begin{align}\label{eqn:doubleanchor}
    &\pest  =\notag\\
    &\hspace{-0em}\frac{(1-g)(1-g)}{(1-g)(1-g) + bg} \PJim ~ \PJip + 
    \notag\\ &~ \hspace{-0em} \frac{(1-g)}{(1-g) + (1-b)} \PJimA \PJip +
    \notag\\ &~ \hspace{-0em} \frac{(1-g)}{(1-g) + (1-b)} \PJim \PJipA +
    \notag\\&~ \hspace{-0em} \frac{bg}{bg+(1-b)(1-b)} \PJimA \PJipA.
\end{align}

One possible drawback of estimating $p_{S_i|Y^n}(J|y^n)$ from $(p_{S_1|Y_1}(J|y_1),\ldots, p_{S_n|Y_n}(J|y_n))$ based on temporal correlation is the risk of increasing the number of false negatives, especially at non-jammed indices neighboring jammed indices. These false negatives could potentially have a negative impact on performance. However, as discussed in Section~\ref{sec:pj:errortypes} and as presented in Section~\ref{sec:res}, their impact on BLER performance is negligible.

\section{Simulation Results}\label{sec:res}

\begin{figure}
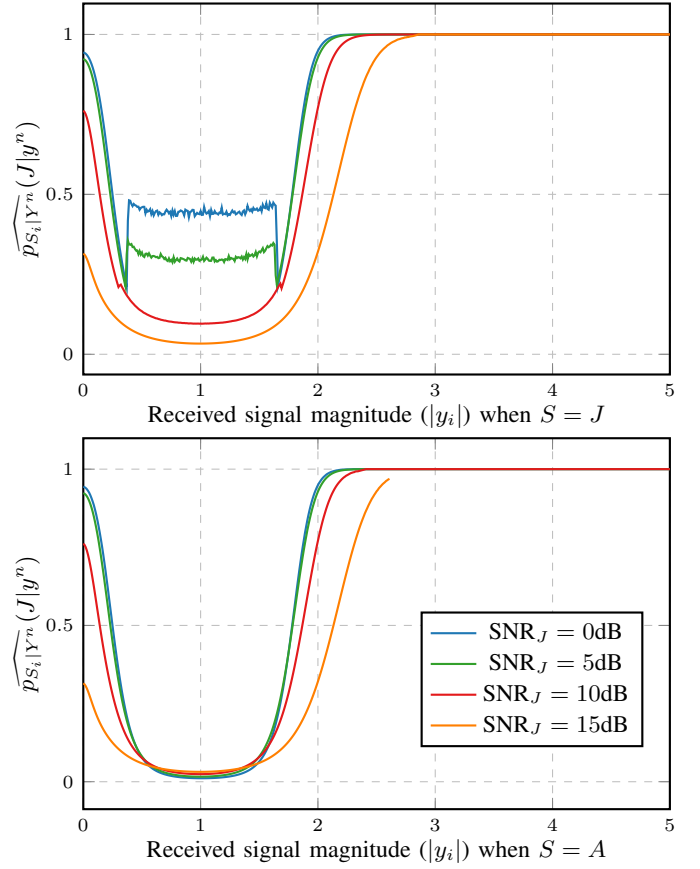

  \centering
  \input{figures/pj_result.tikz}
  \vspace{0.5em}
  \input{figures/pa_result.tikz} 
   \vspace{-1.5em}
   \caption{Simulated $\pest$ based on the received signal magnitude, when $S=J$ (top) and $S=A$ (bottom). The SNR of the AWGN channel is fixed at $\text{SNR}_A=12$ dB. All parameters are kept the same as in Fig.~\ref{fig:pj}.}
  \label{fig:pj2}
  \vspace{-1.0em}
\end{figure}

The proposed jamming-aware LLR calculation using $\pest$ is evaluated. The state transition probabilities are set to $b=0.01$ and $g=0.25$, referring to an overall stationary jamming probability of $\frac{b}{b+g} = 3.84\%$. The SNR for the AWGN state is set as $\text{SNR}_A = 12$ dB. An empirical threshold probability of $0.2$ is used to derive the anchor indices, and the neighboring indices are re-evaluated recursively, \textit{i.e.} until the estimates $\pest$ of $p_{S_i|Y^n}(J|y^n)$ of the neighboring index fall below the threshold.

Fig.~\ref{fig:pj2} visualizes $\pest$ when the ground truth is $S=J$ (top) and $S=A$ (bottom) with respect to the received signal magnitude of an arbitrary index $i$. 
Distinct than Fig.~\ref{fig:pj}, the statistics from states $A$ and $J$ states are kept separate to demonstrate the impact of Markov state transitions. 
All other parameters are kept the same as in Fig.~\ref{fig:pj}. Compared to Fig.~\ref{fig:pj}, the estimate of $p_{S_i|Y^n}(J|y^n)$ near the constellation point has increased significantly for all considered $\text{SNR}_J$ values when $S=J$. This means that the amount of false negatives that originally arise with using (\ref{eqn:pj_init}) solely has decreased significantly. In return, the estimate of $p_{S_i|Y^n}(J|y^n)$ when $S=A$ has not changed significantly compared to the first approximation in Fig.~\ref{fig:pj}. Therefore, false positives due to leveraging temporal correlation with the neighboring indices is negligible. 

\begin{figure}
  \centering
  \scalebox{1.00}{
  \begin{tikzpicture}[spy using outlines=
	{circle, magnification=2.0, connect spies}]
  \pgfplotsset{
    label style = {font=\fontsize{9pt}{7.2}\selectfont},
    tick label style = {font=\fontsize{7pt}{7.2}\selectfont}
  }

\begin{axis}[
	scale = 1,
    ymode=log,
    ylabel={BLER}, ylabel style={yshift=-0.50em},
    xmin = -10,
    xmax = 10,
    ymin = 1e-7,
    ymax = 2,
    grid=both,
    ymajorgrids=true,
    xmajorgrids=true,
    grid style=dashed,
    width=\columnwidth, height=6.25cm,
    thick,
    mark size=3,
]




\addplot[
    color=Paired-1,
    mark=o,
    thick,
    mark size=3,
    mark repeat=3,mark phase=2,
]
table {
-1.000e+01 9.43800e-01
-9.000e+00 9.32800e-01
-8.000e+00 9.16600e-01
-7.000e+00 8.91200e-01
-6.000e+00 8.63100e-01
-5.000e+00 8.19300e-01
-4.000e+00 7.63300e-01
-3.000e+00 6.81900e-01
-2.000e+00 5.83500e-01
-1.000e+00 4.58300e-01
0.000e+00 3.32200e-01
1.000e+00 2.05600e-01
2.000e+00 1.11000e-01
3.000e+00 4.70000e-02
4.000e+00 1.77000e-02
5.000e+00 5.30000e-03
6.000e+00 1.00774e-03
7.000e+00 2.53968e-04
8.000e+00 4.94776e-05
9.000e+00 8.30579e-06
1.000e+01 1.02905e-06
};

\addplot[
    color=Paired-3,
    mark=+,
    thick,
    mark size=3,
    mark repeat=3,mark phase=1,
]
table {
-1.000e+01 2.61700e-01
-9.000e+00 2.58400e-01
-8.000e+00 2.67800e-01
-7.000e+00 2.66900e-01
-6.000e+00 2.59300e-01
-5.000e+00 2.42700e-01
-4.000e+00 2.21100e-01
-3.000e+00 2.01300e-01
-2.000e+00 1.66700e-01
-1.000e+00 1.22600e-01
0.000e+00 8.17000e-02
1.000e+00 4.88000e-02
2.000e+00 2.47000e-02
3.000e+00 9.70000e-03
4.000e+00 3.82643e-03
5.000e+00 1.13861e-03
6.000e+00 3.07074e-04
7.000e+00 7.64584e-05
8.000e+00 1.23663e-05
9.000e+00 2.10000e-06
1.000e+01 3.76087e-07
};

\addplot[
    color=Paired-5,
    mark=x,
    thick,
    mark size=3,
    mark repeat=3,mark phase=3,
]
table {
-1.000e+01 8.96000e-02
-9.000e+00 9.16000e-02
-8.000e+00 1.00100e-01
-7.000e+00 1.02300e-01
-6.000e+00 1.00800e-01
-5.000e+00 9.53000e-02
-4.000e+00 8.79000e-02
-3.000e+00 7.81000e-02
-2.000e+00 6.25000e-02
-1.000e+00 4.66000e-02
0.000e+00 3.04000e-02
1.000e+00 1.79000e-02
2.000e+00 8.90000e-03
3.000e+00 3.86578e-03
4.000e+00 1.47588e-03
5.000e+00 5.31615e-04
6.000e+00 1.64390e-04
7.000e+00 4.11414e-05
8.000e+00 9.42636e-06
9.000e+00 1.67512e-06
1.000e+01 3.66008e-07
};

\end{axis}

\node at (3.70,5.00) {$\text{RLC}[128,105]$};

\end{tikzpicture}}
    \scalebox{1.00}{
  \begin{tikzpicture}[spy using outlines=
	{circle, magnification=2.0, connect spies}]
  \pgfplotsset{
    label style = {font=\fontsize{9pt}{7.2}\selectfont},
    tick label style = {font=\fontsize{7pt}{7.2}\selectfont}
  }

\begin{axis}[
	scale = 1,
    ymode=log,
    xlabel={Jammer SINR (\text{dB})}, xlabel style={yshift=0.4em},
    ylabel={BLER}, ylabel style={yshift=-0.50em},
    xmin = -10,
    xmax = 10,
    ymin = 1e-7,
    ymax = 2,
    grid=both,
    ymajorgrids=true,
    xmajorgrids=true,
    grid style=dashed,
    width=\columnwidth, height=6.25cm,
    thick,
    mark size=3,
    legend cell align={left},
    legend style={at={(0.00,0.14)},anchor=west,
            font=\fontsize{8pt}{7.2}\selectfont},
    legend columns=1,
]




\addplot[
    color=Paired-1,
    mark=o,
    thick,
    mark size=3,
    mark repeat=3,mark phase=2,
]
table {
-1.000e+01 9.45800e-01
-9.000e+00 9.33100e-01
-8.000e+00 9.16000e-01
-7.000e+00 8.93800e-01
-6.000e+00 8.62600e-01
-5.000e+00 8.18000e-01
-4.000e+00 7.63200e-01
-3.000e+00 6.81500e-01
-2.000e+00 5.84700e-01
-1.000e+00 4.57800e-01
0.000e+00 3.24800e-01
1.000e+00 2.05700e-01
2.000e+00 1.09900e-01
3.000e+00 4.77000e-02
4.000e+00 1.71000e-02
5.000e+00 4.32189e-03
6.000e+00 9.50065e-04
7.000e+00 2.50655e-04
8.000e+00 5.87902e-05
9.000e+00 9.55961e-06
1.000e+01 1.49182e-06
};
\addlegendentry{Regular LLR approach}

\addplot[
    color=Paired-3,
    mark=+,
    thick,
    mark size=3,
    mark repeat=3,mark phase=1,
]
table {
-1.000e+01 2.63300e-01
-9.000e+00 2.60300e-01
-8.000e+00 2.62200e-01
-7.000e+00 2.68900e-01
-6.000e+00 2.62200e-01
-5.000e+00 2.48600e-01
-4.000e+00 2.23600e-01
-3.000e+00 1.95800e-01
-2.000e+00 1.59600e-01
-1.000e+00 1.18400e-01
0.000e+00 8.01000e-02
1.000e+00 4.81000e-02
2.000e+00 2.52000e-02
3.000e+00 1.13000e-02
4.000e+00 5.30000e-03
5.000e+00 1.61844e-03
6.000e+00 3.35462e-04
7.000e+00 7.99050e-05
8.000e+00 1.84522e-05
9.000e+00 2.31384e-06
1.000e+01 4.58157e-07
};
\addlegendentry{Proposed LLR w/o Markov state transitions}

\addplot[
    color=Paired-5,
    mark=x,
    thick,
    mark size=3,
    mark repeat=3,mark phase=3,
]
table {
-1.000e+01 9.23000e-02
-9.000e+00 9.19000e-02
-8.000e+00 9.97000e-02
-7.000e+00 1.02400e-01
-6.000e+00 1.01800e-01
-5.000e+00 9.83000e-02
-4.000e+00 8.97000e-02
-3.000e+00 7.91000e-02
-2.000e+00 5.89000e-02
-1.000e+00 4.69000e-02
0.000e+00 2.99000e-02
1.000e+00 1.80000e-02
2.000e+00 8.90000e-03
3.000e+00 4.95835e-03
4.000e+00 1.72753e-03
5.000e+00 6.05525e-04
6.000e+00 1.69870e-04
7.000e+00 4.65500e-05
8.000e+00 1.14828e-05
9.000e+00 1.96254e-06
1.000e+01 3.58873e-07
};
\addlegendentry{Proposed LLR with Markov state transitions}

\end{axis}

\node at (3.70,5.00) {$\text{5G CA-Polar}[128,105]$};
\end{tikzpicture}}
  \caption{BLER comparison of the proposed approach against conventional LLR, with respect to jammer SINR, using $\text{RLC}[128,105]$ (top) and 5G $\text{CA-Polar}[128,105]$ (bottom) codes. The SNR of the AWGN channel is fixed at $\text{SNR}_A = 12$ dB.}
  \label{fig:bler2}
\end{figure}
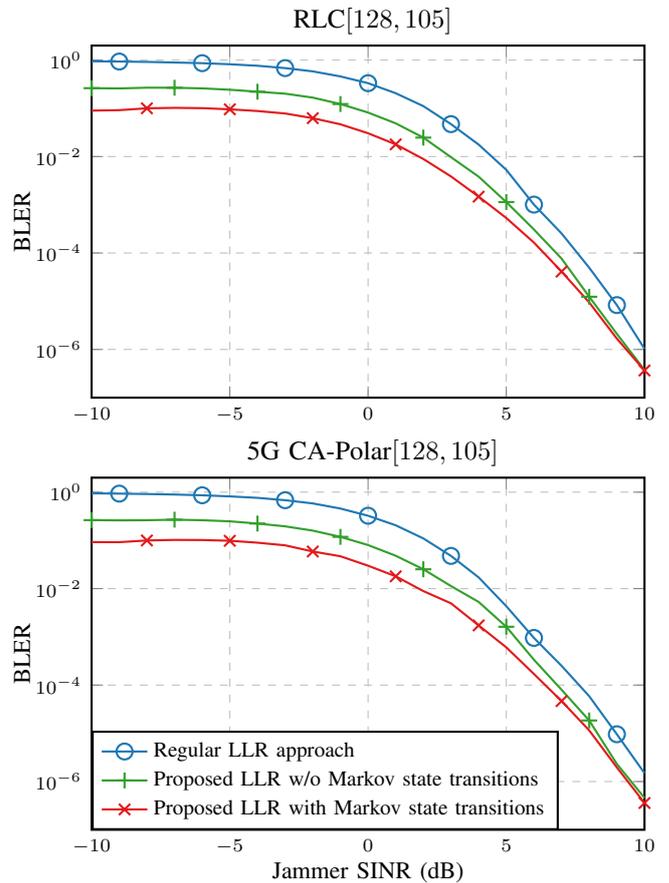

Fig.~\ref{fig:bler2} presents the BLER performance comparison using $\text{RLC}[128,105]$ and 5G NR $\text{CA-Polar}[128,105]$. The ORBGRAND algorithm \cite{duffy2021orbgrand,duffy2022orbgrand} is selected to evaluate the performance of selected codes, since it is a universal soft-information decoder that allows to evaluate distinct codebooks. Moreover, despite its recent introduction to the literature, several works report the practicality of its algorithm family with demonstrated circuit implementations \cite{riaz2021grand,abbas2022orbgrand,condo2022orbgrand}. The jammer SINR represents the legitimate transmission power to the jammer interference power ratio, \textit{i.e.} low SINR indicates a powerful jammer. For both comparison scenarios, the performance using the regular LLR approach (\ref{eqn:llr:awgn}) is the baseline BLER. The red curves represent the proposed approach using $\pest$. The BLER performance for $p_{S_i|Y_i}(J|y_i)$  without using Markov chain state transitions in (\ref{eqn:singleanchor})-(\ref{eqn:doubleanchor}) is also shown as a reference. The baseline performance shows that a strong jammer yields a BLER close to $1$, \textit{i.e.} almost no packets can be decoded, therefore causing a DoS. The proposed LLR computation (\ref{eqn:llr:proposed}) using $\pest$ is shown to improve the baseline BLER performance by an order of magnitude at the DoS region, \textit{i.e.} about $9$ out of $10$ packets can be decoded correctly despite the strong jammer interference. The proposed approach demonstrates $2.7$ dB SINR gain at a BLER of $10^{-2}$ and $0.75$ dB gain at a BLER of $10^{-6}$ for both codes. Note that the high SINR values indicate weak jammers which are not typical since they can only degrade the performance marginally and cannot cause a DoS. Nonetheless, the proposed approach is shown to outperform the baseline even in the high SINR region.

\section{Conclusion}\label{sec:conc}

In this work, a novel and general physical layer security approach against a smart bursty jammer is developed. First, the adversary is modeled as disguised in the channel as a Gaussian variable with zero mean. In addition, the overall active duration for the jammer is determined by a two-state Markov chain with low interference time to avoid RSSI detection. To tackle this challenging model, we proposed a new approach based on LLR calculation under adversarial constraints, to improve the BLER performance. The new LLR calculation is based on a conditional probability of jamming, calculated using the received signal and the Markov chain state transition probabilities. The proposed approach is implemented prior to decoding and works with any soft-information decoder. Simulation results with the universal ORBGRAND algorithm using $\text{RLC}[128,105]$ and 5G $\text{CA-Polar}[128,105]$ codes show that the proposed solution can substantially improves the reliability estimates for the received signals, preventing denial of service, and yields a substantial SNR gain of up to $2.7$ dB. Future work includes further improvement of jamming detection accuracy, and comparing with other available soft-information decoders. 

\section*{Acknowledgements}
This work was partially supported by Defense Advanced
Research Projects Agency Contract number HR00112120008 and by National Science Foundation ECCS Award numbers 2128517 and 2128555. The content of the information does not necessarily reflect the position or the policy of the US Government, and no official endorsement should be inferred. This publication has emanated from research supported in part by a grant from Science Foundation Ireland under grant number 18/CRT/6049. The opinions, findings and conclusions or recommendations expressed in this material are those of the author(s) and do not necessarily reflect the views of the Science Foundation Ireland.

\bibliographystyle{IEEEtran}
\bibliography{IEEEabrv,ref}

\end{document}